\pdfoutput=1
\documentclass[final,5p,times,twocolumn, a4paper]{elsarticle}

\usepackage{atbegshi}
\AtBeginShipout{%
  \ifnum\value{page}=1
    \setbox\AtBeginShipoutBox=\hbox{\hspace*{2mm}\box\AtBeginShipoutBox}%
  \fi
}

\usepackage{amssymb, amsmath, bm, graphicx, subcaption, lineno, mathtools}
\usepackage[numbers]{natbib}  

\graphicspath{{./plots/images}}

\journal{Journal of Computational Physics}

\begin{document}

\begin{frontmatter}

\title{A New Approach to Compute Linear Landau Damping}

\author[ipp,tum]{M. Pelkner\corref{cor1}} 
\author[ipp]{K. Hallatschek} 
\author[ipp]{M. Raeth} 

\cortext[cor1]{Corresponding author. E-mail: maximilian.pelkner@ipp.mpg.de}

\affiliation[ipp]{organization={Max Planck Institute for Plasma Physics},
            addressline={Boltzmannstraße 2}, 
            city={Garching},
            postcode={85748}, 
            country={Germany}}

\affiliation[tum]{organization={TUM School of Natural Sciences},
            addressline={Boltzmannstraße 10}, 
            city={Garching},
            postcode={85748}, 
            country={Germany}}

\begin{abstract}
    \label{sec:abstract}

   We present a semi-analytical method for calculating exact time-domain solutions to linear Landau-damping problems. Unlike traditional residue-based approaches, our framework avoids the need to compute complex zeros of the dielectric function. We illustrate the method for an electrostatic, unmagnetized plasma with kinetic ions and adiabatic electrons by deriving the response of both the ion density and the ion distribution function to an initial density perturbation. We further show that the construction can be formally extended to linear Vlasov-Maxwell problems with a uniform background magnetic field and a Maxwellian equilibrium distribution.

\end{abstract}

\end{frontmatter}

\section{Introduction}
\label{sec:introduction}

\begin{figure}[ht]
    \centering
    \begin{minipage}[t]{0.75\columnwidth}
    \includegraphics[width=\textwidth]{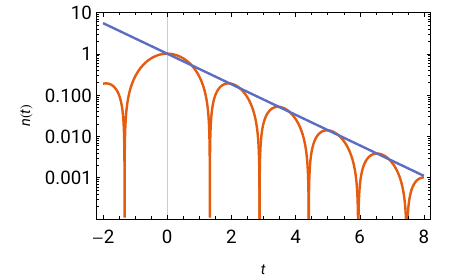}
    \caption{Illustration of the classical Landau damping test \cite{Landau:1946}. The slope of the blue line represents the damping rate prescribed by the dominant Landau pole. The red line depicts the time evolution of the absolute value of an electron density perturbation at a fixed spatial position in the presence of a static ion background. After an initial transient phase, the exponential decay rate converges to the asymptotic value predicted by the dominant Landau pole.}
    \label{fig:standardLandau}
    \end{minipage}
\end{figure}

To ensure the accuracy of simulations in analytically intractable regimes, numerical predictions must be compared against available exact solutions whenever possible. However, exact solutions of the Vlasov equation are notoriously difficult to derive, and benchmarks are therefore often restricted to indirect diagnostics or simplified limits. A classical example is the Landau damping test, which measures the decay rate of electron Langmuir oscillations in the presence of a static ion background (Fig. \ref{fig:standardLandau}) \cite{Kormann:2019, Finn:2023}. Such residue-based analyses are most effective in the late-time regime, where the solution is governed by a small number of dominant poles of the dielectric function. Resolving the transient response near the initial time, however, requires an increasingly large number of poles. Determining these poles numerically can be challenging and, depending on the background distribution function, may even be impossible if the dielectric function cannot be analytically continued sufficiently far into the lower complex half-plane. In this work, we present a semi-analytical method that circumvents pole summation entirely.

The remainder of this paper is organized as follows. First, we rederive the exact solution of the linearized Vlasov equation in Fourier space under the assumptions of quasi-neutrality, an unmagnetized plasma, and adiabatic electrons. These assumptions are motivated by the current configuration of the BSL6D code \cite{Kormann:2019,Schild:2024} and serve to simplify the initial presentation. We then introduce the semi-analytical method by calculating the ion density response to an initial ion density perturbation. Building on this result, we determine the response of the linearized ion distribution function, which requires treating the resonant singularity on the real axis associated with the continuous spectrum of Case-van Kampen modes. Finally, we provide a formal argument showing that the extension to magnetized configurations follows the same logic, provided that the background magnetic field is uniform and the equilibrium distribution is Maxwellian. The explicit construction in this setting involves significant additional algebraic complexity and is therefore left for future work.

\section{Analytic distribution-function response}
\label{sec:analytical distribution function response}

In the following, we assume that collisions can be neglected, magnetic fields are absent, and the distribution function $f(x,v,t)$ of a particle species can be separated into an equilibrium part $f_0(v)$ and a small perturbation $f_1(x,v,t)$,
\begin{equation}
    f(x,v,t)= f_0(v) + f_1(x,v,t) .
\end{equation}
Our goal is to calculate the time evolution of $f_1$ from an arbitrary initial condition $f_1(x,v,t=0)$, such as the pure density perturbation considered in the next section. The stationary background is assumed to be of the form
\begin{equation}
    \label{bgDistribution}
    f_0(v) = n_0 f_{\text{M}}(v),
\end{equation}
where $n_0$ is the constant equilibrium density and $f_{\text{M}}(v)$ is the Maxwellian distribution,
\begin{equation}
    \label{maxwell}
    f_{\text{M}}(v) \coloneq \frac{1}{\sqrt{2 \pi} v_{\text{th}}} e^{-\frac{v^2}{2 v_{\text{th}}^2}}, \qquad v_{\text{th}} \coloneq \sqrt{\frac{T}{m}}.
\end{equation}
Here, $T$ denotes the temperature. Since we restrict our analysis to small perturbations $f_1$, nonlinear terms can be neglected, and the perturbation is governed by the linearized Vlasov equation \cite{Goldston:2019},
\begin{equation}
    \label{linearVlasov3D}
    \partial_t f_1 + \bm{v} \cdot \nabla f_1 - \frac{q}{m} \nabla \phi \cdot \nabla_{\bm{v}} f_0 = 0.
\end{equation}
Because the homogeneous, unmagnetized background distribution \eqref{bgDistribution} is translationally invariant and Eq.~\eqref{linearVlasov3D} is linear, spatial Fourier modes evolve independently. For each wave vector, we may rotate the coordinate system such that the $x$-axis is aligned with the wave vector. Denoting the corresponding coordinate and velocity component by $x$ and $v$, respectively, we can therefore restrict the following discussion to the representative one-dimensional problem
\begin{equation}
    \label{linearVlasov}
    \partial_t f_1 + v \partial_x f_1 - \frac{q}{m} \partial_x \phi \partial_v f_0 = 0 .
\end{equation}
The electrostatic potential $\phi$ is determined by imposing quasineutrality, $n_{\text{e}} = n_{\text{i}}$, together with an adiabatic electron response \cite{Knorr:1970},
\begin{equation}
    \label{phiAdiabaticElectrons}
    \phi(x,t) \coloneq \frac{T}{q n_0} \int dv\, f_1(x,v,t) = \frac{T}{q n_0} n_1(x,t).
\end{equation}
Thus, $\phi$ is first order in the perturbation amplitude. To construct a solution of the linearized Vlasov equation \eqref{linearVlasov} that satisfies a given initial condition, $f_1(x,v,t=0)$, we follow Case \cite{Case:1959} and introduce
\begin{equation}
    \label{posSol}
    f_1^+(x,v,t)=\theta(t)f_1(x,v,t) ,
\end{equation}
where the Heaviside step function is defined as
\begin{equation}
    \theta(t) \coloneq 
\begin{cases} 
0 & t < 0 \\
1 & t \geq 0 
\end{cases} .
\end{equation}
To transform $f_1^+$ to the Fourier domain, we use the Fourier transform convention
\begin{equation}
    \label{FT}
    \hat{g}(k, \omega) \coloneq \frac{1}{2 \pi} \int_{-\infty}^{\infty} dx dt g(x, t) e^{- i(k x - \omega t)} ,
\end{equation}
where $\hat{g}$ denotes the Fourier transform of $g$. Inserting $f_1^+$ into Eq.~\eqref{linearVlasov}, using the convention \eqref{FT}, and assuming $\operatorname{Im}(\omega)>0$ to ensure convergence of the Fourier integral of $f_1^+$, we obtain
\begin{align}
    \label{linearVlasovFourier}
    - i\omega \hat{f}_1^+(k,v,\omega) + &i k v \hat{f}_1^+(k,v,\omega) - \frac{q}{m} i k \hat{\phi}^+(k,\omega) \partial_v f_0 \notag \\
    & \qquad \qquad \qquad = \frac{1}{\sqrt{2 \pi}} \hat{f}_1(k,v,t=0) .
\end{align}
In a slight departure from the notation specified in Eq.~\eqref{FT}, $\hat{f}_1(k,v,t=0)$ refers specifically to the spatial Fourier transform of the initial condition. For the following calculations, it is convenient to treat $\omega$ as a real variable. Since $\operatorname{Im}(\omega)>0$, we introduce a small positive parameter $\epsilon > 0$ and perform the substitution $\omega \rightarrow \omega + i \epsilon$, $\omega \in \mathbb{R}$. Solving Eq.~\eqref{linearVlasovFourier} for $\hat{f}_1^+$ then yields 
\begin{align}
    \label{f1evolution}
    \hat{f}_{1,\epsilon}^+ (k,v,\omega) \coloneq& \frac{k v}{\omega - k v + i \epsilon} f_{\text{M}}(v) \frac{q n_0}{T} \hat{\phi}_{1, \epsilon}^+(k,\omega) \notag \\
    &+ \frac{1}{\sqrt{2 \pi}}\frac{i}{\omega - k v + i \epsilon} \hat{f}_1(k,v,t=0) ,
\end{align}
where, according to Eq.~\eqref{phiAdiabaticElectrons},
\begin{equation}
    \frac{q n_0}{T} \hat{\phi}_{1, \epsilon}^+(k,\omega) = \int dv \hat{f}_{1,\epsilon}^+ (k,v,\omega) .
\end{equation}

\section{Semi-analytical density response}
\label{sec:semi-analytical density response}

\begin{figure}[h]
    \centering
    \begin{minipage}[t]{0.75\columnwidth}
        \centering
        \includegraphics[width=\textwidth]{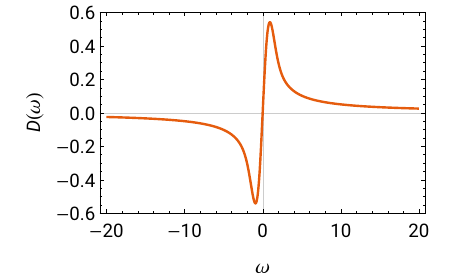}
        \caption{The Dawson function along the real $\omega$-axis. For all $\zeta \in \mathbb{C}$, $D(\zeta)$ is entire, antisymmetric and asymptotically proportional to $1/\zeta$ as $|\zeta| \rightarrow \infty$.}
    \label{fig:DawsonF}
    \vspace{0.5cm}
    \end{minipage}
    \begin{minipage}[t]{0.75\columnwidth}
        \centering
        \includegraphics[width=\textwidth]{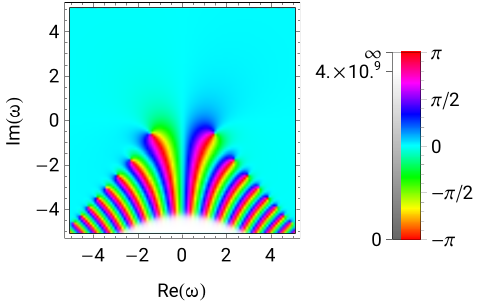}
        \caption{Complex plot of the denominator in Eq.~\eqref{badBacktrafo}. Zeros are identified by a $2\pi$ phase winding around closed contours. The phase is color-coded using a rainbow spectrum.}
    \label{fig:complexRoots}
    \end{minipage}
\end{figure}

We now assume a single-mode density perturbation at $t=0$, i.e., 
\begin{equation}
    \label{initialCondition}
   f_1(x,v,t=0) = f_{\text{M}}(v) e^{i k x}, 
\end{equation}
and compute the density response for $t>0$. General initial perturbations will be discussed in the next section. For the perturbation \eqref{initialCondition}, Eq.~\eqref{linearVlasovFourier} admits a plane-wave solution,
\begin{equation}
    \label{f1p}
   \hat{f}_1^+(k,v,\omega) \coloneq \hat{f}_1^+(v,\omega) e^{i k x} ,
\end{equation}
which consequently imposes a plane-wave structure on $\hat{\phi}$ according to Eq.~\eqref{phiAdiabaticElectrons},
\begin{equation}
    \hat{\phi}(k,\omega) = \frac{T}{q n_0} \int dv \hat{f}_1^+(v,\omega) e^{i k x} =  \frac{T}{q n_0} \hat{n}_1^+(\omega) e^{i k x} .
\end{equation}
To keep the notation concise, we use the same label for $\hat{f}_1(k,v,\omega)$ and $\hat{f}_1(v,\omega)$ and set $k=1$, since the solution for an arbitrary $k$ can be recovered from the self-similarity of the distribution-function response \eqref{f1evolution}. Integrating Eq.~\eqref{f1evolution} with respect to velocity yields
\begin{align}
    \label{n1evolution}
    \hat{n}^+_{1,\epsilon}(\omega) =& - \hat{n}^+_{1,\epsilon} (\omega) \int_{-\infty}^{\infty} dv \frac{v f_{\text{M}}(v)}{v - (\omega + i\epsilon)} \notag \\
    &- \frac{i}{\sqrt{2 \pi}} \int_{-\infty}^{\infty} dv \frac{f_{\text{M}}(v)}{v - (\omega + i\epsilon)} .
\end{align}
By expressing the velocity integrals in terms of the plasma dispersion function $Z(\zeta)$ \cite{NRL},
\begin{equation}
    \label{zFunctionA}
    Z(\zeta) \coloneq \frac{1}{\sqrt{\pi}} \int_{-\infty}^{\infty} dx \frac{e^{-x^2}}{x - \zeta} \qquad \text{for} \qquad \operatorname{Im}(\zeta) > 0 ,
\end{equation}
and nondimensionalizing with $m = T = 1$, we obtain
\begin{equation}
    \label{n1evolution2}
    \hat{n}^+_{1,\epsilon}(\omega) = - \frac{i}{2\sqrt{\pi}} \frac{Z(\xi)}{2 + \xi Z(\xi)} , \qquad \text{where} \qquad \xi \coloneq \frac{\omega + i \epsilon}{\sqrt{2}} .
\end{equation}
As written, Eq.~\eqref{n1evolution2} is defined only in the upper complex half-plane owing to the definition \eqref{zFunctionA}. Consequently, evaluating the $t>0$ density response by Fourier inversion, 
\begin{equation}
    \label{badBacktrafo}
\hat{n}_1^+ (t) = - \frac{i}{\sqrt{2 \pi}} \int_{-\infty}^{\infty} \frac{d \omega}{\sqrt{2 \pi}} \frac{Z(\omega)}{2 + \omega Z(\omega)} e^{-i \sqrt{2} \omega t},
\end{equation}
requires the use of the analytic continuation of the plasma dispersion function \cite{Fried:1961},
\begin{equation}
    \label{zFunction}
    Z(\zeta) \coloneq i \sqrt{\pi} e^{-\zeta^2} - 2 D(\zeta), \qquad \zeta \in \mathbb{C},
\end{equation}
where
\begin{equation}
    \label{dawsonFunction}
    D(\zeta) \coloneq e^{-\zeta^2} \int_0^{\zeta} dx e^{x^2}.
\end{equation}
Equation~\eqref{zFunction} defines an entire function; Fig.~\ref{fig:DawsonF} shows the Dawson function $D(\zeta)$ along the real axis. In the numerical evaluations below, we use the implementation of $D(\zeta)$ provided by Wolfram Mathematica \cite{dawsonf}. 

\begin{figure}[h]
    \centering
    \begin{minipage}[t]{0.75\columnwidth}
        \centering
        \includegraphics[width=\textwidth]{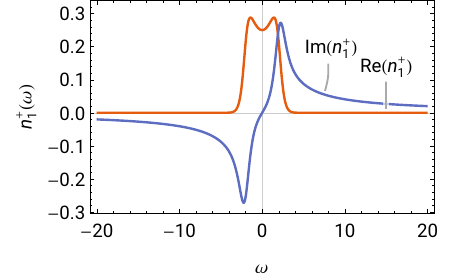} 
    \caption{Fourier spectrum of the density response $n_1^+(\omega)$.}
    \label{fig:spectrumBadBacktrafo}
        \vspace{0.5cm}
    \end{minipage}
    \begin{minipage}[t]{0.75\columnwidth}
        \centering
        \includegraphics[width=\textwidth]{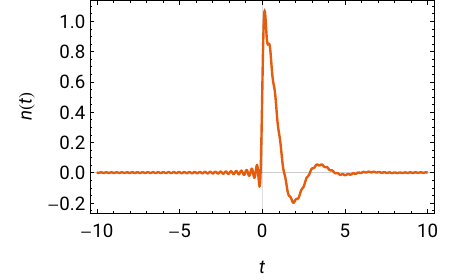}
        \caption{Numerical Fourier inversion of the density response $n_1^+$, evaluated over the domain $\omega \in [-20,20]$. At $t=0$, the solution converges to half of the exact initial value.}
        \label{fig:w20}
    \end{minipage}
\end{figure}

Since $Z$ is entire and the zeros of the denominator in \eqref{badBacktrafo} are confined to the lower complex half-plane (Fig.~\ref{fig:complexRoots}), the integrand in \eqref{badBacktrafo} is free of singularities along the integration contour. However, decomposing $\hat{n}^+_1$ into its real and imaginary components along the real $\omega$ line,
\begin{align}
    \label{n1EvolutionDawson}
    \hat{n}_1^+(\omega) =& \frac{2 e^{-\xi^2}}{\pi \xi^2 e^{-2\xi^2} 
    +  4(1 - \xi D(\xi))^2} \notag \\
    &+ \frac{i}{\sqrt{\pi}} \frac{- \pi \xi e^{-2 \xi^2} + 4 D(\xi) \left(1 - \xi D(\xi)\right)}{\pi \xi^2 e^{-2\xi^2} +  4(1 - \xi D(\xi))^2} ,
\end{align}
shows that the real part is symmetric and exhibits Gaussian decay in $\omega$, whereas the imaginary part is antisymmetric and decays only as $\omega^{-1}$ (Fig.~\ref{fig:spectrumBadBacktrafo}). The slow decay of the imaginary component leads to poor convergence of the Fourier integral \eqref{badBacktrafo}; consequently, truncating the integration domain induces spurious Gibbs oscillations (Fig.~\ref{fig:w20}). 

To eliminate this algebraic tail, we symmetrize the density response in Fourier space,
\begin{equation}
    \hat{n}_1^{\text{sym}} (\omega) \coloneq \hat{n}_1^+(\omega) + \hat{n}_1^+(-\omega).
\end{equation}
In the time domain, this corresponds to replacing the one-sided response by its time-symmetric extension, thereby removing the jump at $t=0$. The $\omega^{-1}$ tail is a direct consequence of this jump, since the density response,
\begin{equation}
    n_1^+(x,t) = \int dv\, f_1^+(x,v,t),
\end{equation}
inherits the causal step at the initial time, and the Fourier transform of a step function,
\begin{equation}
    \label{fourierHeaviside}
    \int_{-\infty}^{\infty} \frac{d t}{\sqrt{2 \pi}} e^{i \omega t} \theta(t)
    = \operatorname{p.v.} \left(\frac{i}{\sqrt{2 \pi}\omega}\right)
    + \sqrt{\frac{\pi}{2}}\delta(\omega),
\end{equation}
contains a principal-value contribution proportional to $\omega^{-1}$ \cite{Kanwal:2004}. Using Eq.~\eqref{n1EvolutionDawson}, we obtain
\begin{equation}
    \label{n1EvolutionDawsonSym}
    \hat{n}_1^{\text{sym}} (\omega)
    = \frac{2 e^{-\xi^2}}{\pi \xi^2 e^{-2\xi^2} +  4(1 - \xi D(\xi))^2},
    \qquad
    \xi = \frac{\omega}{\sqrt{2}},
\end{equation}
and define
\begin{equation}
    \label{n1SymResp}
    \hat{n}_1^{\text{sym}}(t) \coloneq
    \int_0^\infty d\omega\, \cos(\omega t)\, \hat{n}_1^{\text{sym}}(\omega).
\end{equation}
\begin{figure}[ht]
    \centering
    \begin{minipage}[t]{0.75\columnwidth}
        \centering
        \includegraphics[width=\textwidth]{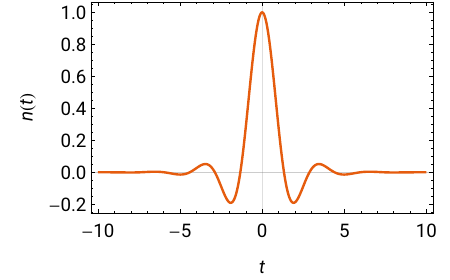}
        \caption{Numerical Fourier inversion of the symmetrized density response $n_1^{\text{sym}}$, evaluated over the domain $\omega \in [0,4]$. The initial condition is accurately recovered.}
        \label{fig:correctSolution}
    \end{minipage}
\end{figure}
Physically, the symmetrized density response $n_1^{\text{sym}}(t)$ 
(Fig.~\ref{fig:correctSolution}) corresponds to a specially prepared scenario 
in which the distribution function evolves from a structured perturbation at 
$t=-\infty$ into a pure density perturbation at $t=0$. For $t>0$, this 
perturbation subsequently decays through phase mixing. It may appear that other 
choices of the $t<0$ continuation of $n_1$ could also remove the spectral tail, 
provided that their evolution from $t=-\infty$ yields the prescribed initial 
condition at $t=0$. However, this is not the case: the symmetry of the initial 
condition, together with the time-reversal invariance of the underlying dynamics, determines a unique $t<0$ solution. We elaborate on this point in the next section.

\begin{figure}[ht]
    \centering
    \begin{minipage}[t]{0.75\columnwidth}
        \centering
        \includegraphics[width=\textwidth]{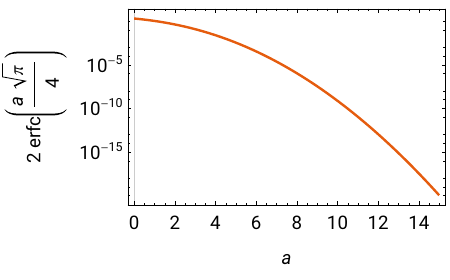}
        \caption{Logarithmic plot of the error bound \eqref{errorBound} as a function of the integration cutoff frequency $a$.}
    \label{fig:erfc}
    \vspace{0.5cm}
    \end{minipage}
    \begin{minipage}[t]{0.75\columnwidth}
        \centering
        \includegraphics[width=\textwidth]{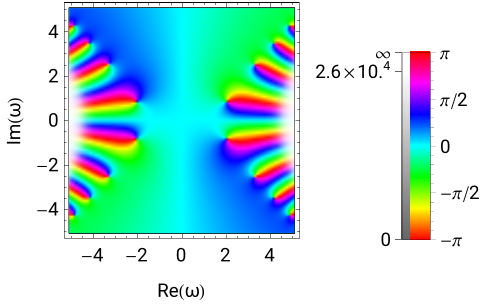}
        \caption{Complex plot of the denominator of the symmetrized density response $n_1^{\text{sym}}$, similar to Fig.~\ref{fig:complexRoots}. Because of the symmetrization, the zeros of the denominator, corresponding to poles of the spectrum, are now located in both the upper and lower complex half-planes.}
        \label{fig:roots_sym}
    \end{minipage}
\end{figure}

The two principal sources of numerical error are the truncation error due to the finite integration domain and the quadrature error introduced by discretizing the Fourier integral. To derive an error bound for the former, we neglect the quadrature error and consider the exact integral in Eq.~\eqref{n1SymResp}. For the spectrum considered here, we use the empirically verified bound
\begin{equation}
    \hat{n}_1^{\text{sym}}(\omega) < e^{-\frac{\pi \omega^2}{16}} \qquad \forall \omega .
\end{equation}
For the truncation error $E_a(t)$,
\begin{equation}
   E_a(t) \coloneq n_1(t) - \int_0^a d \omega \cos(\omega t) \hat{n}_1^{\text{sym}}(\omega),
\end{equation}
we obtain (Fig.~\ref{fig:erfc})
\begin{equation}
    \label{errorBound}
    |E_a(t)| \leq \int_a^\infty d \omega e^{-\frac{\pi \omega^2}{16}} 
    = 2 \operatorname{erfc} \left ( \frac{a \sqrt{\pi}}{4} \right ).
\end{equation}
To estimate the quadrature error, we first note that the spectrum \eqref{n1EvolutionDawsonSym} is analytic in a strip of width $2c$ around the real axis, where $c$ is the minimum distance from the dominant Landau pole to the real axis (Fig.~\ref{fig:roots_sym}). For our choice of parameters, $c \approx 0.85$. Since the integrand is analytic in this strip, the numerical integration can be performed efficiently using the trapezoidal rule on a uniform grid with spacing $h$. Within the regime of analyticity, the trapezoidal rule exhibits spectral convergence of the quadrature error $ |E_h|$ \cite{trefethen:2014}, i.e.,
\begin{equation}
    |E_h| \leq \frac{2 M}{e^{2 \pi c/h}-1}, \quad \text{with} \quad M \coloneq \sup_{\left |s \right | < c} \int _{-\infty}^{\infty} \left |\hat{n}_1^{\text{sym}}(\omega + i s)\right | d \omega .
\end{equation}
Consequently, the quadrature error can be readily reduced below the level of the truncation error, leaving the latter as the dominant source of numerical uncertainty.

\section{Semi-analytical distribution-function response}
\label{sec:semi-analytical distribution function response}

Any solution of the homogeneous linearized Vlasov equation \eqref{linearVlasov} admits the decomposition 
\begin{equation}
    \label{homSolSeparated}
    f_1(x,v,t) = f_1^+(x,v,t) + f_1^-(x,v,t) ,
\end{equation}
where $f_1^+$ is defined by Eq.~\eqref{posSol} and $f_1^-$ is defined by
\begin{equation}
    \label{negSol}
    f_1^-(x,v,t) \coloneq \theta(-t) f_1(x,v,t) .
\end{equation}
While $\hat{f}_1^+$ solves Eq.~\eqref{linearVlasovFourier} for $\operatorname{Im}(\omega)>0$, 
\begin{equation}
    \label{linearVlasovFourierRep}
    (- i\omega + i k v )\hat{f}_1^+(k,v,\omega) - \frac{q}{m} i k \hat{\phi}^+(k,\omega) \partial_v f_0 = \frac{1}{\sqrt{2 \pi}} \hat{f}_1(k,v,t=0) ,
\end{equation}
$\hat{f}_1^-$ solves
\begin{equation}
    \label{linearVlasovFourierNeg}
    (- i\omega + i k v ) \hat{f}_1^-(k,v,\omega) - \frac{q}{m} i k \hat{\phi}^-(k,\omega) \partial_v f_0 = - \frac{1}{\sqrt{2 \pi}} \hat{f}_1(k,v,t=0) 
\end{equation}
for $\operatorname{Im}(\omega) < 0$. As can be seen from Eq.~\eqref{f1evolution}, the source terms on the right-hand side of Eqs.~\eqref{linearVlasovFourierRep} and \eqref{linearVlasovFourierNeg} induce a slow $\sim \omega^{-1}$ decay in the individual spectra of $\hat{f}_1^+$ and $\hat{f}_1^-$. In the combined spectrum $\hat{f}_1 = \hat{f}_1^+ + \hat{f}_1^-$, however, these terms cancel. Thus, provided that the imposed initial condition admits a smooth solution $f_1$ with compact support, the combined spectrum decays faster than any inverse power of $\omega$ \cite{Stein:2011}. This rapid decay ensures that truncation errors in the Fourier inversion become negligible outside a finite integration domain.

For the special case of a time-symmetric initial condition, time-reversing the governing differential equation for $\hat{f}_1^+$, Eq.~\eqref{linearVlasovFourierRep}, and comparing the result with Eq.~\eqref{linearVlasovFourierNeg} yields
\begin{equation}
    \label{symmterizationCondition}
    \hat{f}_1^-(v,\omega) = \hat{f}_1^+(-v,-\omega) \quad \text{if} \quad \hat{\phi}^+(-\omega) = \hat{\phi}^-(\omega) .
\end{equation}
Thus, according to Eqs.~\eqref{phiAdiabaticElectrons} and \eqref{homSolSeparated}, $\hat{f}_1$ is constructed by time-symmetrizing the $\hat{f}_1^+$ solution, in agreement with the results of the previous section. Analogously, for a time-antisymmetric initial condition satisfying
\begin{equation}
    f_1(x,v,t=0) = - f_1(x,-v,t=0),
\end{equation}
one finds
\begin{equation}
    \label{antisymmetrizationCondition}
    \hat{f}_1^-(v,\omega) = - \hat{f}_1^+(-v,-\omega) \quad \text{if} \quad \hat{\phi}^+(-\omega) = - \hat{\phi}^-(\omega) ,
\end{equation}
i.e., $\hat{f}_1$ is constructed by time-antisymmetrizing the $\hat{f}_1^+$ solution. Arbitrary initial conditions may be analyzed by decomposing the initial state into symmetric and antisymmetric components and treating each contribution as outlined above. Alternatively, one may compute $\hat{f}_1^+$ and $\hat{f}_1^-$ independently.

\begin{figure}[ht]
    \centering
    \begin{minipage}[t]{0.75\columnwidth}
        \centering
        \includegraphics[width=\textwidth]{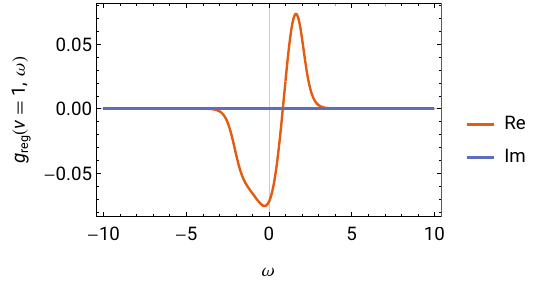}
        \caption{Spectrum of the regularized principal-value integrand \eqref{principalValueRegularized}.}
        \label{fig:pvregSpectrum}
    \end{minipage}
\end{figure}

Returning to the time-symmetric case, we recall Eq.~\eqref{f1evolution},
\begin{align}
    \label{f1evolution2}
\hat{f}_{1,\epsilon}^+ (k,v,\omega) \coloneq& 
\frac{k v}{\omega - k v + i \epsilon} f_{\text{M}}(v) 
\hat{n}_{1,\epsilon}^+(\omega) \notag \\
&+ \frac{1}{\sqrt{2 \pi}}
\frac{i}{\omega - k v + i \epsilon} \hat{f}_1(k,v,t=0) ,
\end{align}
where $\omega,\epsilon \in \mathbb{R}$ and $\epsilon > 0$. Symmetrizing this expression in $\omega$ yields
\begin{align}
    \label{f1Sym}
    \hat{f}_{1,\epsilon}^{\text{sym}} (v,\omega) \coloneq& 
    \frac{\omega - v}{(\omega - v)^2 + \epsilon^2} 
    v f_{\text{M}}(v) \hat{n}_1^{\text{sym}}(\omega) + \notag \\
    &\frac{\epsilon}{(\omega - v)^2 + \epsilon^2} 
    \left(\sqrt{\frac{2}{\pi}} n_0 f_{\text{M}}(v) 
    - i v f_{\text{M}}(v) \hat{n}_1^{\text{asym}}(\omega)\right) ,
\end{align}
with $\hat{n}_1^{\text{asym}}(\omega) \coloneq 
\hat{n}_1^+(\omega)-\hat{n}_1^+(-\omega)$. Taking the limit $\epsilon \to 0$ and applying the Sokhotski-Plemelj theorem gives the distributional limit
\begin{align}
    \label{f1evolution1}
    \lim_{\epsilon \to 0} \hat{f}_{1,\epsilon}^{\text{sym}} (v,\omega) =\: 
    & \operatorname{p.v.} \left[
    \frac{1}{\omega - v} v f_{\text{M}}(v) 
    \hat{n}_1^{\text{sym}}(\omega)\right] + \notag \\
    &\pi \delta(\omega - v) 
    \left(\sqrt{\frac{2}{\pi}} n_0 f_{\text{M}}(v) 
    - i v f_{\text{M}}(v) \hat{n}_1^{\text{asym}}(\omega)\right), 
\end{align}
where $\operatorname{p.v.}$ denotes the Cauchy principal value. Equation~\eqref{f1evolution1} identifies the Case--van Kampen eigenmodes, represented by the delta distributions, as the continuous spectrum of the linearized Vlasov operator \cite{Case:1959, VanKampen:1955}. In particular, the damped Landau modes are not part of this spectrum; rather, they emerge from a superposition of these eigenmodes. To compute the Fourier inversion of the symmetrized distribution-function response,
\begin{align}
    \label{symf1resp}
    f_1^{\text{sym}} (v,t) &= \int \frac{d \omega}{\sqrt{2 \pi}} 
    \hat{f}_1^{\text{sym}}(v,\omega) e^{-i \omega t} \notag \\
    &\coloneq \lim_{\epsilon \to 0} \int \frac{d \omega}{\sqrt{2 \pi}}  
    \hat{f}_{1,\epsilon}^{\text{sym}} (v,\omega)e^{-i \omega t},
\end{align}
we evaluate the delta-function contribution in Eq.~\eqref{f1evolution1} analytically and compute the remaining principal-value term,
\begin{equation}
    \label{principalValue}
    \operatorname{p.v.} \int \frac{d \omega}{\sqrt{2 \pi}} 
    \frac{1}{\omega - v} v f_{\text{M}}(v) 
    \hat{n}_1^{\text{sym}}(\omega) e^{-i \omega t},
\end{equation}
numerically. Apart from the resonant singularity at $\omega=v$, the integrand in Eq.~\eqref{principalValue} is well behaved because $\hat{n}_1^{\text{sym}}$ decays rapidly. For numerical integration along the real $\omega$-axis, we subtract the singular part by introducing the regularized integrand (Fig.~\ref{fig:pvregSpectrum})
\begin{equation}
    \label{principalValueRegularized}
    \hat{g}_{\text{reg}}(v,\omega) \coloneq 
    \frac{1}{\omega - v} v f_{\text{M}}(v) \hat{n}_1^{\text{sym}}(\omega) 
    - c(v) \frac{e^{-(\omega - v)^2}}{\omega - v} ,
\end{equation}
where
\begin{equation}
    c(v) \coloneq v f_{\text{M}}(v) \hat{n}_1^{\text{sym}}(v).
\end{equation}
The exponential factor in Eq.~\eqref{principalValueRegularized} is crucial: it removes the pole locally without introducing a spurious $\omega^{-1}$ tail.

To recover the full solution, we first Fourier-invert the subtracted singular term,
\begin{equation}
    \label{poleFourier}
    \hat{p}(v,\omega) \coloneq c(v) \frac{e^{-(v-\omega)^2}}{\omega - v} ,
\end{equation}
analytically \cite{Kanwal:2004},
\begin{equation}
    \label{pole}
    p(v,t) = i c(v) \sqrt{\frac{\pi}{2}}\operatorname{erf} \bigg(\frac{t}{2} \bigg) e^{-i v t} , 
\end{equation}
where
\begin{equation}
    \label{erfDefinition}
    \operatorname{erf}(\zeta) \coloneq \frac{2}{\sqrt{\pi}} \int_0^{\zeta} e^{-x^2}\,dx, \qquad \zeta \in \mathbb{C}.
\end{equation}
The full time evolution of the distribution function is then obtained by combining the numerical integral of the regularized integrand with the analytically evaluated terms,
\begin{align}
    \label{f1SymEvolution}
    f^{\text{sym}}_1(v,t) =& \int \frac{d\omega}{\sqrt{2 \pi}}\hat{g}_{\text{reg}}(v,\omega) e^{-i \omega t} + p(v,t) \notag \\
    &+ f_{\text{M}}(v) \bigg(1 - i v \sqrt{\frac{\pi}{2}} \hat{n}_1^{\text{asym}}(v) \bigg) e^{-i v t} . 
\end{align}

\begin{figure}[ht]
    \centering
    \begin{minipage}[t]{0.75\columnwidth}
        \centering
        \includegraphics[width=\textwidth]{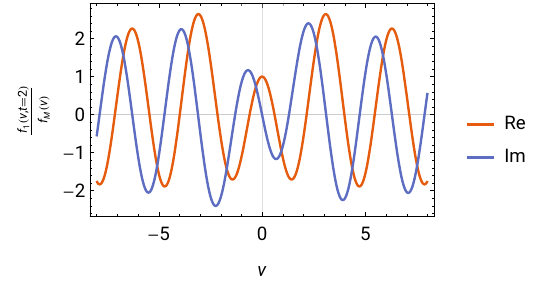}
        \caption{Distribution-function perturbation $f_1$ at fixed time, normalized by a Maxwellian to remove the Gaussian suppression in $v$. As expected from time-reversal symmetry, the real part is symmetric in $v$.}
        \label{fig:f1_pert_to_free_streaming_fixed_t}
        \vspace{0.5cm}
    \end{minipage}
    \begin{minipage}[t]{0.65\columnwidth}
        \centering
        \includegraphics[width=\textwidth]{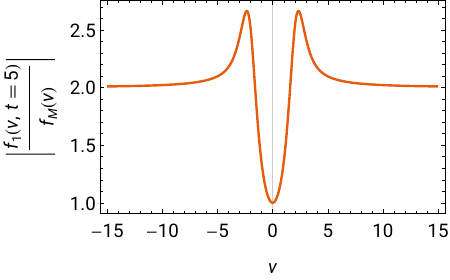}
        \caption{Absolute value of $f_1$ at fixed time, normalized by the background Maxwellian. The time is chosen such that the Landau-damped terms are sufficiently suppressed.}
        \label{fig:abs_value_asymptotics}
    \end{minipage}
\end{figure}

We conclude this section by discussing the constant-time and constant-velocity behavior of Eq.~\eqref{f1SymEvolution}. At fixed time, all terms in Eq.~\eqref{f1SymEvolution} are suppressed by at least a Gaussian factor in velocity. Hence, the perturbation $f_1$ vanishes at large velocities for any fixed time (Figs.~\ref{fig:f1_pert_to_free_streaming_fixed_t} and \ref{fig:abs_value_asymptotics}). 

At fixed velocity, Eq.~\eqref{f1SymEvolution} is more naturally analyzed by decomposing the perturbation $f_1$ into an adiabatic part and a non-adiabatic perturbation $h$,
\begin{equation}
    \label{f1decomp}
    f_1(x,v,t) = h(x,v,t) - \frac{q n_0}{T} \phi(x,t) f_{\text{M}}(v) .
\end{equation}
With the adiabatic-electron closure, the time evolution of the non-adiabatic perturbation $h$ is governed by the inhomogeneous equation
\begin{equation}
    \label{nonAdPert}
    \partial_t h + v \partial_x h = f_0 \partial_t n_1 .
\end{equation}

\begin{figure}[ht]
    \centering
    \begin{minipage}[t]{0.75\columnwidth}
        \centering
        \includegraphics[width=\textwidth]{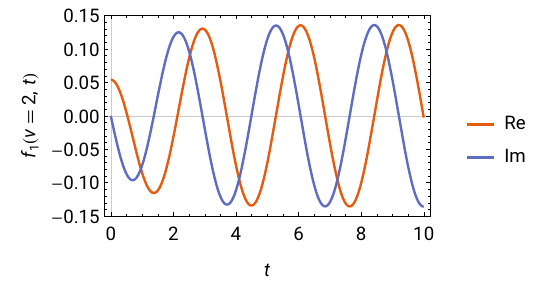}
        \caption{Distribution-function perturbation at fixed velocity. The solution transitions from a perturbed state with amplitude modulations to the asymptotic free-streaming solution.}
        \label{fig:f1_pert_to_free_streaming_fixed_v}
    \end{minipage}
\end{figure}

A general solution of Eq.~\eqref{nonAdPert} consists of a homogeneous and a particular contribution, $h = h_{\text{hom}} + h_{\text{part}}$, where $h_{\text{hom}}$ satisfies the free-streaming equation
\begin{equation}
    \label{nonAdPertHom}
    \partial_t h_{\text{hom}} + v \partial_x h_{\text{hom}} = 0 .
\end{equation}
Since $\phi \to 0$ and $\partial_t n_1 \to 0$ as $t \to \infty$, only the homogeneous contribution persists asymptotically. Thus, at fixed velocity, $f_1$ approaches the free-streaming solution after the Landau-damped contribution has decayed (Fig.~\ref{fig:f1_pert_to_free_streaming_fixed_v}).

\begin{figure}[ht]
    \centering
    \begin{minipage}[t]{0.75\columnwidth}
        \centering
        \includegraphics[width=\textwidth]{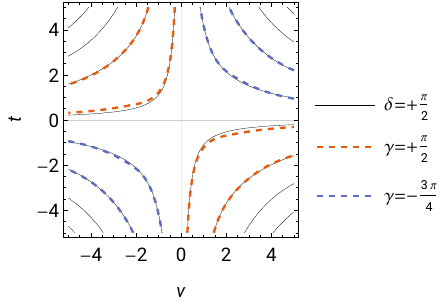}
        \caption{Contour plot in the $t-v$ plane. Solid black lines indicate contours of the exact complex argument $\delta$ of the distribution-function response $f_1$. The dashed red and blue lines represent lines of constant $\gamma$.}
        \label{fig:comp_phase_asymptotics}
    \end{minipage}
\end{figure}

\begin{figure}[ht]
    \centering
    \begin{minipage}[t]{0.75\columnwidth}
        \centering
        \includegraphics[width=\textwidth]{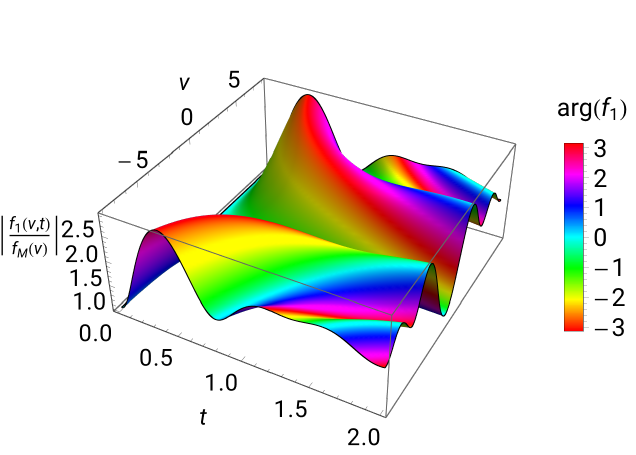}
        \caption{Normalized absolute value of $f_1$ for $t \geq 0$ and a fixed wavenumber ($k=1$). The complex argument $\delta$ is indicated by a rainbow color map.}
        \label{fig:f1_small}
    \end{minipage}
    \vspace{0.5cm}
    \begin{minipage}[t]{0.75\columnwidth}
        \centering
        \includegraphics[width=\textwidth]{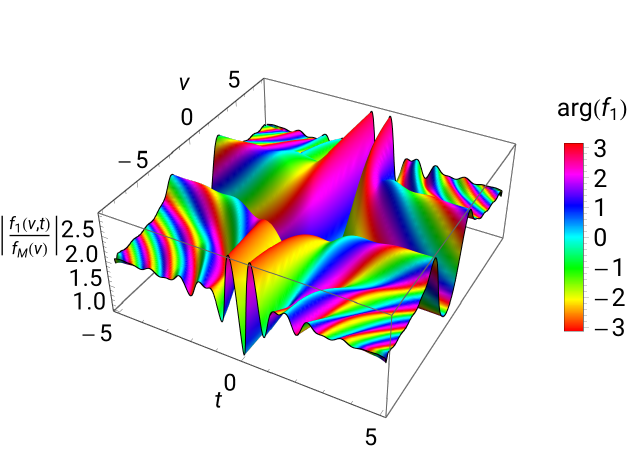}
        \caption{Plot analogous to Fig.~\ref{fig:f1_small}, but for a larger section of the $t-v$ plane.}
        \label{fig:f1_large}
    \end{minipage}
\end{figure}

To illustrate the asymptotic value of $h_{\text{hom}}$, we normalize Eq.~\eqref{f1SymEvolution} by a Maxwellian and evaluate it at a time $\tau$ for which the Landau-damped terms are sufficiently suppressed and $\operatorname{erf}(\tau/2) \approx 1$. This gives the asymptotic estimate
\begin{equation}
    \label{absPlot}
     |f_1(v,\tau)| \approx \left | i c(v) \sqrt{\frac{\pi}{2}} e^{-i v \tau} + f_{\text{M}}(v) \left (1 - i v \sqrt{\frac{\pi}{2}}\hat{n}_1^{\text{asym}}(v) \right ) e^{-i v \tau} \right | ,
\end{equation}
as illustrated in Figs.~\ref{fig:abs_value_asymptotics}, \ref{fig:f1_small}, and \ref{fig:f1_large}. The asymptotic phase is obtained analogously from
\begin{equation}
    \label{phasePlot}
    \gamma \coloneq \operatorname{arg} \left [ i c(v) \sqrt{\frac{\pi}{2}} e^{-i v t} + f_{\text{M}}(v) \left (1 - i v \sqrt{\frac{\pi}{2}}\hat{n}_1^{\text{asym}}(v) \right ) e^{-i v t} \right ] ,
\end{equation}
as shown in Fig.~\ref{fig:comp_phase_asymptotics}. Figures~\ref{fig:f1_small} and \ref{fig:f1_large} provide a combined visualization of the normalized amplitude $|f_1|/f_{\text{M}}$ and the complex argument $\delta \coloneq \operatorname{arg}(f_1)$. 

\begin{figure}[h]
    \centering
    \begin{minipage}[t]{0.75\columnwidth}
        \centering
        \includegraphics[width=\textwidth]{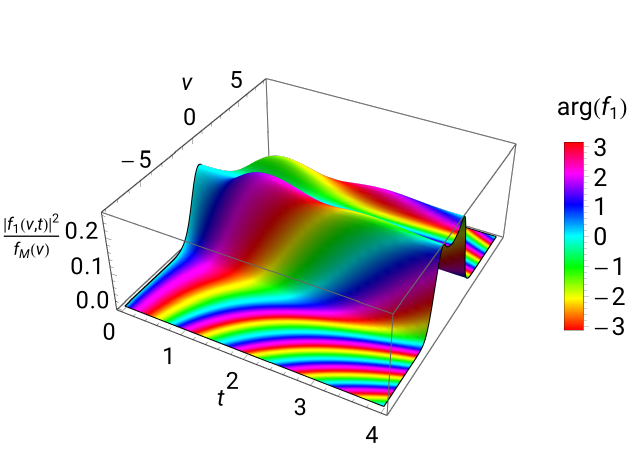}
        \caption{Ion free-energy density for a fixed wavenumber ($k=1$). As in Figs.~\ref{fig:f1_small} and \ref{fig:f1_large}, the complex argument $\delta$ is indicated by a rainbow color map.}
        \label{fig:free_energy_density}
        \vspace{0.5cm}
    \end{minipage}
    \begin{minipage}[t]{0.75\columnwidth}
        \centering
        \includegraphics[width=\textwidth]{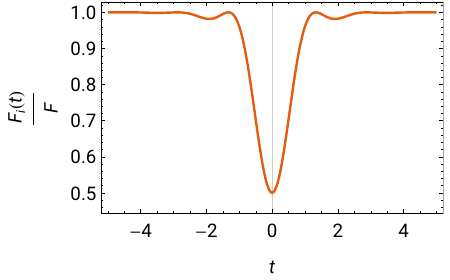}
        \caption{Relative ion free energy $F_{\text{i}}(t)/F$ for a fixed wavenumber ($k=1$). At $t=0$, the ions carry exactly half of the free energy, indicating that the remaining energy resides in the adiabatic electron response.}
        \label{fig:free_energy}
    \end{minipage}
\end{figure}

Finally, we verify the physical consistency of the solution by examining the ion free-energy density (Fig.~\ref{fig:free_energy_density}) and the corresponding ion free energy (Fig.~\ref{fig:free_energy}). In the absence of external sources, such as temperature gradients, the self-consistent evolution of $f_1$ is constrained by the total free energy of the system,
\begin{equation}
    \label{freeEnergy}
    F \coloneq \sum_s \int d v \frac{(f_1^s)^2}{2 f^s_0},
\end{equation}
where the sum runs over the particle species \cite{Goldston:2019}. For the time-symmetric initial condition \eqref{initialCondition}, the symmetrized solution $f_1^{\text{sym}}(t)$ in Eq.~\eqref{symf1resp} describes a scenario in which the ions gradually transfer free energy to the initially unperturbed adiabatic electron response over the interval $t \in (-\infty,0]$. At $t=0$, both species carry equal amounts of free energy, and the electron perturbation reaches its maximum. For $t>0$, the electron response decays by Landau damping, returning the free energy to the ions. 

\section{Formal generalization}
\label{sec:generalization}

We finally discuss under which conditions the outlined construction of a numerically well-behaved response can be applied to general initial conditions within the linearized Vlasov-Maxwell framework. The derivation of more general response functions beyond those presented so far is left for future work; our current objective is to show that the numerical method itself is, in principle, extensible.

To this end, the magnetic field is decomposed into a zeroth-order contribution, representing a strong external background field, and a small first-order perturbation,
\begin{equation}
    \bm{B} \coloneq  \bm{B}_{\text{ext}} + \bm{B}_{\text{pert}} .
\end{equation}
We assume that $\bm{B}_{\text{ext}}$ is time-independent and transforms as $\bm{B}_{\text{ext}} \rightarrow - \bm{B}_{\text{ext}}$ under time reversal, consistent with $\bm{v} \rightarrow - \bm{v}$. Moreover, as before, we assume a Maxwellian equilibrium distribution,
\begin{equation}
    \label{3dMaxwellian}
    f_0^s(\bm{v}) = n_{0s} \left ( \frac{m_s}{2 \pi T_s} \right )^{\frac{3}{2}} e^{-\frac{m_s v^2}{2 T_s}} ,
\end{equation}
where $s$ denotes the species index. The linearized, three-dimensional Vlasov equation then takes the form
\begin{equation}
    \label{linearVlasovGeneral}
    \partial_t f^s_1 + \bm{v} \cdot \nabla f^s_1 + \bm{F} \cdot \nabla_{\bm{v}} f^s_0 + 
    \left(\bm{v} \times \bm{B}_{\text{ext}}\right) \cdot \nabla_{\bm{v}} f^s_1 = 0 ,
\end{equation}
and the perturbing force reads
\begin{equation}
    \label{perturbationForce}
    \bm{F} \coloneq \frac{q_s}{m_s} \left ( \bm{E} + \bm{v} \times \bm{B}_{\text{pert}} \right ) .
\end{equation}
In Eq.~\eqref{linearVlasovGeneral}, we have absorbed the $q_s/m_s$ prefactor into the definition of the external magnetic field. The self-consistent fields are defined by
\begin{equation}
    \bm{E} = - \nabla \phi - \partial_t \bm{A} \quad \text{and} \quad 
    \bm{B}_{\text{pert}} = \nabla \times \bm{A} .
\end{equation}
In the Lorenz gauge, the equations governing $\phi$ and $\bm{A}$ are \cite{Jackson}
\begin{equation}
    \label{maxwellEqnsLorentzGauge}
    \left ( \Delta -\frac{1}{c^2} \partial_t^2 \right ) \phi = - \frac{\rho}{\epsilon_0} 
    \quad \text{and} \quad 
    \left ( \Delta -\frac{1}{c^2} \partial_t^2 \right ) \bm{A} = - \mu_0 \bm{j} ,
\end{equation}
where
\begin{equation}
    \rho \coloneq \sum_s q_s\int d^3 v\, f^s_1 
    \quad \text{and} \quad 
    \bm{j} \coloneq \sum_s q_s \int d^3 v\, \bm{v} f^s_1 .
\end{equation}
We now set $c=1$ and proceed analogously to Section~\ref{sec:analytical distribution function response}. For $\operatorname{Im}(\omega)>0$, the positive-time solution $\hat{f}_1^{s+}$ is determined by
\begin{align}
    \label{retSolf}
   \left ( - i \omega - i \bm{k} \cdot \bm{v} + \bm{v} \times \bm{B}_{\text{ext}}\cdot \nabla_{\bm{v}} \right )\hat{f}_1^{s +} + &\hat{\bm{F}}^+(\bm{k}, \omega) \cdot \nabla_{\bm{v}} f_0^s \notag \\
   & \qquad =  \hat{f}_1(\bm{k},\bm{v}, t=0),
\end{align}
whereas for $\operatorname{Im}(\omega)<0$, the negative-time solution $\hat{f}_1^{s-}$ satisfies
\begin{align}
    \label{advSolf}
   \left ( - i \omega - i \bm{k} \cdot \bm{v} + \bm{v} \times \bm{B}_{\text{ext}} \cdot \nabla_{\bm{v}} \right )\hat{f}^{s-}_1 + &\hat{\bm{F}}^-(\bm{k}, \omega) \cdot \nabla_{\bm{v}} f_0^s \notag \\
   & \quad = - \hat{f}_1(\bm{k},\bm{v}, t=0) .
\end{align}
Here, prefactors of $(2 \pi)^{-3/2}$ from the spatial Fourier transform have been absorbed into the initial conditions. Equations~\eqref{retSolf} and \eqref{advSolf} implicitly require spatial homogeneity of both the magnetic background and the equilibrium distributions, since any spatial dependence would introduce convolutions in Fourier space instead of multiplications, rendering the semi-analytical treatment analytically intractable.

Analogously to $f_1^+$ and $f_1^-$, the scalar and vector potentials are decomposed as
\begin{equation}
    \label{MaxwellFieldPos}
    \phi^+ \coloneq \theta(t) \phi, \quad 
    \bm{A}^+ \coloneq \theta(t) \bm{A},
\end{equation}
and
\begin{equation}
    \label{MaxwellFieldNeg}
    \phi^- \coloneq \theta(-t) \phi, \quad 
    \bm{A}^- \coloneq \theta(-t) \bm{A} .
\end{equation}
To transform the positive-time fields into the frequency domain and obtain an expression for $\hat{\bm{F}}^+$, we assume $\operatorname{Im}(\omega)>0$ and first apply the Fourier transform \eqref{FT} to the second time derivative of $\phi^+$,
\begin{align}
&\int_{-\infty}^{\infty} \frac{dt}{\sqrt{2 \pi}} e^{i \omega t} \partial_t^2 \phi^+ \notag \\
& \quad = \int_0^{\infty} \frac{dt}{\sqrt{2 \pi}} e^{i \omega t} \partial_t^2 \phi \notag \\
& \quad = \frac{1}{\sqrt{2 \pi}}\left [ e^{i\omega t} \partial_t \phi \right ]_0^{\infty} 
- \int_{-\infty}^{\infty} \frac{dt}{\sqrt{2 \pi}} (i \omega) e^{i \omega t}\partial_t \phi^+ \notag \\
& \quad = - \frac{1}{\sqrt{2 \pi}}\dot{\phi}(0) 
- \frac{i \omega}{\sqrt{2 \pi}} \left [ e^{i\omega t} \phi \right ]_0^{\infty} 
- \omega^2 \int_{-\infty}^{\infty} \frac{dt}{\sqrt{2 \pi}} e^{i \omega t} \phi^+ \notag \\
& \quad =  \frac{1}{\sqrt{2 \pi}} \left(- \dot{\phi}(0) + i \omega \phi(0)\right) - \omega^2 \hat{\phi}^+ .
\end{align}
Proceeding analogously for $\bm{A}^+$, we obtain from Eq.~\eqref{maxwellEqnsLorentzGauge}
\begin{align}
    \label{retSol}
    \hat{\bm{A}}^+(\bm{k}, \omega) 
    &= \frac{1}{k^2 - \omega^2} 
    \left [ - \mu_0 \hat{\bm{j}}^+_1(\bm{k},\omega) 
    - \frac{1}{\sqrt{2 \pi}} \left (\dot{\bm{A}}(0) - i \omega \bm{A}(0)\right )\right ], \notag \\ 
    \hat{\phi}^+(\bm{k}, \omega) 
    &= \frac{1}{k^2 - \omega^2} 
    \left [ \frac{\hat{\rho}^+_1(\bm{k},\omega)}{\epsilon_0} 
    - \frac{1}{\sqrt{2 \pi}} \left (\dot{\phi}(0) - i \omega \phi(0)\right )\right ] .
\end{align}
Conversely, for the advanced fields we assume $\operatorname{Im}(\omega)<0$ and obtain from Eq.~\eqref{MaxwellFieldNeg}
\begin{align}
    \label{advSol}
    \hat{\bm{A}}^-(\bm{k}, \omega) 
    &= \frac{1}{k^2 - \omega^2} 
    \left [ - \mu_0 \hat{\bm{j}}^-_1(\bm{k},\omega) 
    + \frac{1}{\sqrt{2 \pi}} \left (\dot{\bm{A}}(0) - i \omega \bm{A}(0)\right )\right ], \notag \\ 
    \hat{\phi}^-(\bm{k}, \omega) 
    &= \frac{1}{k^2 - \omega^2} 
    \left [ \frac{\hat{\rho}^-_1(\bm{k},\omega)}{\epsilon_0} 
    + \frac{1}{\sqrt{2 \pi}} \left (\dot{\phi}(0) - i \omega \phi(0)\right )\right ] .
\end{align}

Adding $f_1^{s+}$ and $f_1^{s-}$ now cancels the source terms in Eqs.~\eqref{retSolf} and \eqref{advSolf}. Likewise, the initial-field terms in Eqs.~\eqref{retSol} and \eqref{advSol} cancel upon summation. Consequently, provided that the initial states $f_1^s(t=0)$ are consistent with a smooth evolution of the coupled system, the resulting spectrum again decays faster than any inverse power of $\omega$ \cite{Stein:2011}. Whether a symmetrization similar to that presented in Section~\ref{sec:analytical distribution function response} is possible, however, depends crucially on the boundary conditions chosen for the electric and magnetic potentials.

The construction admits the following physical interpretation. Once the initial conditions for the distribution functions and the fields at $t=0$ are specified, a unique distribution-function perturbation is prepared at $t=-\infty$, together with a self-consistent radiation field. Owing to the use of advanced potentials for $t<0$, the plasma absorbs this radiation over the interval $t \in (-\infty,0]$ until the prescribed state at $t=0$ is reached. For $t>0$, the perturbation decays through standard Landau damping, and the radiation field is re-emitted into the environment. The resulting formulation is manifestly time-symmetric.

\section{Conclusion}
\label{sec:conclusion}

In this work, we have shown that exact time-domain solutions of linear Landau-damping problems can be obtained by numerical Fourier inversion without determining the complex zeros of dielectric functions. The central idea of the method is to construct the distribution-function perturbation $f_1$ as a superposition of the forward-evolving solution $f_1^+$ and the backward-evolving solution $f_1^-$, both of which are uniquely determined by the initial condition $f_1(x,v,t=0)$. Equivalently, the construction can be viewed as implicitly defining a prehistory of the distribution function that evolves from $t=-\infty$ toward the prescribed initial condition at $t=0$, thereby removing the discontinuity associated with a one-sided response. Provided that the initial condition admits a smooth time evolution and that $f_1$ remains compactly supported, the spectrum of $f_1$ decays superalgebraically. Spectral contributions outside a finite frequency interval are therefore negligible.

\begin{figure}[ht]
    \centering
    \includegraphics[scale=0.85]{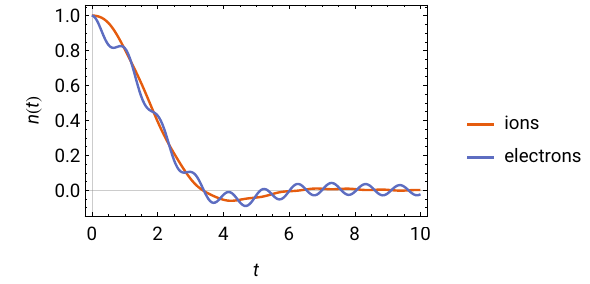} 
    \caption{Density response of a two-species electron-ion plasma with self-consistent Poisson coupling. The parameters $m_e/m_i=1/10$, $\omega_{\text{isw}}=\sqrt{T_i/m_i}=0.5$, $\omega_{\text{pe}}=10 \omega_{\text{isw}}$, $n_0^i=n_0^e=1$, and $T_e/T_i=1$ were chosen for illustrative purposes.}
    \label{fig:langmuir}
\end{figure}

As a concrete application, we calculated the responses of the ion density and the ion distribution function to an initial density perturbation in an unmagnetized plasma with kinetic ions and adiabatic electrons. An illustrative generalization is the self-consistent evolution of an unmagnetized two-species Vlasov-Poisson plasma. This model captures the coupling between high-frequency electron Langmuir waves and low-frequency ion-acoustic oscillations; the resulting density response is shown in Fig.~\ref{fig:langmuir}.

\begin{figure}[ht]
    \centering
    \includegraphics{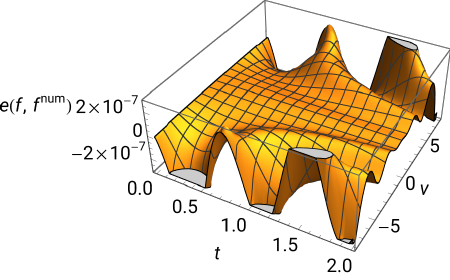} 
    \caption{Pointwise relative error $e \left (f,f^{\text{num}} \right )$ of a BSL6D distribution-function simulation. BSL6D is a semi-Lagrangian, full-$f$ Vlasov ion code with adiabatic electrons that employs Strang splitting in time and Lagrange interpolation in space to update the distribution function at discrete grid points by tracing characteristics backward in phase space \cite{Kormann:2019, Schild:2024}. For this plot, a 1D1V simulation with $32 \times 65$ grid points on the domain $[0,2\pi]\times[-8,8]$ was performed over the time interval $[0,2]$ with $\Delta t = 0.01$, using an eight-point stencil for the Lagrange interpolation. In BSL6D, time is measured in units of inverse ion gyrofrequencies. The relative error increases toward the edges of the velocity domain because the Lagrange interpolation error depends on the advection shift. For larger velocities, this shift is larger for a given time step, resulting in a higher interpolation error.}
    \label{fig:bsl6d_comp}
\end{figure}

As a final benchmark, we compare the distribution-function response derived in Sec.~\ref{sec:semi-analytical distribution function response} with a fully kinetic 1D1V simulation. Figure~\ref{fig:bsl6d_comp} shows the pointwise relative error between the semi-analytical distribution function, Eq.~\eqref{f1SymEvolution}, and the numerical solution, defined by
\begin{equation}
    e \left (f,f^{\text{num}} \right ) = \frac{f^{\text{num}}(k=1,v,t)}{f_{\text{M}}^{\text{num}}(v)} \frac{f_{\text{M}}(v)}{f(k=1,v,t) + 1} - 1 .
\end{equation}
The superscript $\mathrm{num}$ denotes quantities obtained from the numerical simulation, and the relative error is evaluated at the fixed Fourier mode $k=1$. Because the initial perturbation is sinusoidal, this corresponds to comparing a single spatial Fourier component, with the initial perturbation amplitude normalized to unity.

\section{Acknowledgments}
\label{sec:acknowledgments}

    This work has been carried out partly within the framework of the EUROfusion
    Consortium, funded by the European Union via the Euratom Research and Training
    Programme (Grant Agreement No 101052200 – EUROfusion). Support has also been
    received by the EUROfusion High Performance Computer (Marconi-Fusion). Views and
    opinions expressed are however those of the author(s) only and do not
    necessarily reflect those of the European Union or the European Commission.
    Neither the European Union nor the European Commission can be held responsible
    for them.  Numerical simulations were performed at the MARCONI-Fusion
    supercomputer at CINECA, Italy, and at the HPC system at the Max Planck
    Computing and Data Facility (MPCDF), Germany.

\footnotesize
\bibliographystyle{elsarticle-num}      
\bibliography{references}               

@book{Goldston:2019,
	author = {Goldston R. J; Rutherford P. H},
	title = {Introduction to plasma physics},
	publisher = {Taylor \& Francis Groups},
	year = {2019},
}

@Article{Landau:1946,
  author  = {Landau, L. D.},
  journal = {J. Phys. (USSR)},
  title   = {On the vibrations of the electronic plasma},
  year    = {1946},
  pages   = {25--34},
  volume  = {10},
}

@article{VanKampen:1955,
title = {On the theory of stationary waves in plasmas},
journal = {Physica},
volume = {21},
number = {6},
pages = {949-963},
year = {1955},
author = {N.G. {Van Kampen}},
}

@article{Case:1959,
title = {Plasma oscillations},
journal = {Annals of Physics},
volume = {7},
number = {3},
pages = {349-364},
year = {1959},
author = {K.M Case},
}

@Book{NRL,
  author    = {Richardson, A.S. and United States. Office of Naval Research and Naval Research Laboratory (U.S.)},
  publisher = {Naval Research Laboratory},
  title     = {2019 NRL Plasma Formulary},
  year      = {2019},
}

@book{Stein:2011,
  title={Fourier Analysis: An Introduction},
  author={Stein, E.M. and Shakarchi, R.},
  series={Princeton lectures in analysis},
  year={2011},
  publisher={Princeton University Press}
}

@article{Knorr:1970,
year = {1970},
volume = {12},
number = {12},
pages = {927},
author = {G Knorr and J Nuehrenberg},
title = {The adiabatic electron plasma and its equation of state},
journal = {Plasma Physics},
}

@Book{Fried:1961,
  author    = {Fried, B.D. and Conte, S.D.},
  publisher = {Academic Press},
  title     = {The Plasma Dispersion Function: The Hilbert Transform of the Gaussian},
  year      = {1961},
}

@book{Kanwal:2004,
author = {Kanwal, Ram P.},
publisher = {Birkhäuser Boston},
title = {Generalized Functions : Theory and Applications },
year = {2004},
}

@Article{Kormann:2019,
  author    = {Kormann, K. and Reuter, K. and Rampp, M.},
  journal = {The International Journal of High Performance Computing Applications},
  title   = {A massively parallel semi-Lagrangian solver for the six-dimensional Vlasov–Poisson equation},
  year    = {2019},
  pages   = {924-947},
  volume  = {33},
}

@Article{Finn:2023,
author = {Finn, D. and Knepley, M. and Pusztay, J. and Adams, M.},
journal = {Communications in Applied Mathematics and Computer Science},
year = {2023},
title = {A Numerical Study of Landau Damping with PETSc-PIC},
pages = {135-152},
volume = {18}
}

@article{Schild:2024,
    title = {A performance portable implementation of the semi-Lagrangian algorithm in six dimensions},
    author = {Nils Schild and Mario Räth and Sebastian Eibl and Klaus Hallatschek and Katharina Kormann},
    journal = {Computer Physics Communications},
    volume = {295},
    pages = {108973},
    year = {2024},
}

@misc{dawsonf,
  author = {{Wolfram Research, Inc.}},
  title = {DawsonF},
  howpublished = {Wolfram Language Online Reference Guide},
  year = {1988}
}

@book{Jackson,
      author        = "Jackson, John David",
      title         = "{Classical electrodynamics}",
      publisher     = "Wiley",
      address       = "New York, NY",
      year          = "1975"
}

@article{trefethen:2014,
  author  = {Trefethen, Lloyd N. and Weideman, J. A. C.},
  title   = {The Exponentially Convergent Trapezoidal Rule},
  journal = {SIAM Review},
  volume  = {56},
  number  = {3},
  pages   = {385-458},
  year    = {2014},
  publisher = {SIAM}
}

\end{document}